\documentclass[12pt,english,twoside]{article}

\usepackage{AK}
\usepackage{amsfonts}
\usepackage{amssymb}
\usepackage{amsmath}
\usepackage{newlfont}
\usepackage[longnamesfirst,sort]{natbib}
\usepackage[svgnames]{xcolor}
\usepackage{graphicx}
\usepackage[colorlinks=true,bookmarks=false,implicit=false,urlcolor=darkblue]{hyperref}
\usepackage{qrcode}

\bibpunct{(}{)}{,}{a}{,}{,}
\newcommand{\SR}{{\rm I\kern-.21em R}}

\newcommand{\eenn}{\end{eqnarray}}

\newcommand{\qrendproof}{\hspace{0.25em}{\raisebox{0.1em}{\bf{{\qrcode[height=0.6em]{Q.E.D.}}}}}}

\def\ds{\displaystyle}

\def\cR{{\cal R}}

\def\player(#1,#2){\langle#1,#2\rangle}

\newcommand\blfootnote[1]{%
  \begingroup
  \renewcommand\thefootnote{}\footnote{#1}%
  \addtocounter{footnote}{-1}%
  \endgroup
}
\input tcilatex
\begin{document}
\definecolor{darkblue}{rgb}{0,0,0.35} 
\title{
 \fontsize{32}{24}\selectfont{Tournament Auctions$^{*}$}}
\author{\hspace{-2em}\href{http://www.anderlini.net}{Luca Anderlini}\\
\hspace{-2em}\begin{tabular}{c}\textit{\href{http://www.georgetown.edu}{Georgetown University} and}\\
\textit{\href{http://www.unina.it/en_GB/home}{University of Naples Federico II}}
\end{tabular}
\and \href{https://uk.linkedin.com/in/gaon-kim-53a636194?}{GaOn Kim}\\
\textit{\href{http://www.romemaster.it}{RoME} --- \href{https://www.eief.it/eief/}{EIEF}}
\\
 \vspace{0.45em}
}

\date{\vspace{3em}\data}

\maketitle
\keywords{Tournament Auctions, Overbidding, Revenue Equivalence\hspace{-0.3em}
\vspace{0.2em}
 } 

\jel{\href{http://www.aeaweb.org/econlit/jelCodes.php?view=econlit}{C70},
\href{http://www.aeaweb.org/econlit/jelCodes.php?view=econlit}{C72},
\href{http://www.aeaweb.org/econlit/jelCodes.php?view=econlit}{C79.}$\,$\hspace{-0.8em}
\vspace{0.2em}
 } 

\simplecorrespondent{\href{http://anderlini.net}{Luca Anderlini} $\;$ --- $\;$
\href{mailto:luca@anderlini.net}{{\tt luca@anderlini.net}}}

\runninghead{\textsc{Tournament Auctions}}

\runningauthor{\textsc{Anderlini and Kim}} 

\definecolor{darkblue}{rgb}{0,0,0.35} %
\vspace{3em}

\begin{abstract}\hspace{-1.1em}We examine ``tournament'' second-price auctions in which $N$ bidders compete
for the right to participate in a second stage and contend against bidder $N+1$. 

When the first $N$ bidders are committed so that their bids cannot be changed in the second 
stage, the analysis yields some unexpected results. 

The first $N$ bidders consistently bid
above their values in equilibrium. When bidder $N+1$ is sufficiently stronger
than the first $N$, overbidding leads to an increase in expected revenue 
in comparison to the standard second-price auction when $N$ is large.

\blfootnote{$^{*}$ Both authors are grateful to EIEF in Rome for hospitality and support.}
\end{abstract}

\section{Introduction}\label{Section: Introduction}
We examine ``tournament'' second-price auctions in which $N$ bidders compete
for the right to participate in a second stage and contend against bidder $N+1$. 

When the first $N$
bidders can submit fresh bids in the second stage this is equivalent to a second price auction.
When the first $N$ bidders are committed so that their bids cannot be changed in the second 
stage, they
consistently bid {\it above} their values in equilibrium. 

When all bidders are ex-ante identical 
overbidding results in a decrease in expected revenue for the auctioneer. 
If instead bidder $N+1$ is sufficiently stronger than the first $N$ 
overbidding {\it increases} expected
revenue relative to the standard second-price auction when $N$ is 
large.

All proofs are relegated to an Appendix.
In the numbering of equations, definitions, lemmas etc prefix of ``A'' indicates that the relevant
item is to be found in the Appendix.

\section{Related Literature}\label{Section: related Literature}

Even attempting a review of the literature on auctions in a short note would be foolish. Given that
one of our central results concerns expected revenue citing \citet{Vickrey:61} cannot be avoided.
The rest we leave to a survey by \citet{Klemperer:99}, to the colossal edited collection of papers
found in \citet{Klemperer:00}, and a favorite textbook by \citet{Krishna:09}.

\section{Setup}\label{Section: Setup}

\subsection{General}\label{Subsection: General}
A single indivisible object is for sale, and there are 
$N \geq 2$ first-stage bidders and $1$ second-stage bidder. 
First-stage bidders 
have independent private values $v_i \overset{\mathrm{iid}}{\sim} F$ 
with support $[0,\bar{v}]$ 
with $\bar{v}>0$. The distribution $F$ is 
absolutely continuous with density $f(v)$ $>$ $0$
$\forall v \in [0,\bar{v}]$. For simplicity, $f(\cdot)$ is also 
assumed to be bounded and 
differentiable.

The second-stage bidder $N+1$ has value $w \sim G$ with 
support $[0,\bar{w}]$ with $\bar{w}$ $\geq$ $\bar{v}$. 
The distribution $G$ is absolutely continuous with density 
$g(w)$ $>$ $0$ $\forall w \in [0,\bar{w}]$.
For simplicity $g(\cdot)$ is also assumed to be bounded and 
differentiable.
The value
$w$ is independent of the values $v_i$ of the first stage bidders.

Bids in both first and second stage are sealed.
Bidder $N+1$ competes with the highest bidder of 
the first stage.\footnote{It seems more appealing to assume, 
as we do, that
$N+1$ does not observe the actual first stage winning bid 
although this is completely irrelevant. All our
results holds unchanged if $N+1$ observes the actual 
bid of the winner of the first stage.}

Some significant details of the two-stage auction will change as 
we consider different cases. 
In all cases, the values of all $N+1$ bidders are private 
information to each of the
participants and are independently drawn {\it once and for all} 
at the beginning of the game.

Throughout, by ``equilibrium'' we mean a
Perfect Bayesian Equilibrium of the two-stage incomplete information game at hand. 
We also restrict attention to equilibria with
bidding functions that are
non-decreasing in values and identical across the $N$ first-stage 
bidders.\footnote{\label{footnote: ties broken randomly}Any 
ties are broken randomly,
with all tied bids winning with strictly positive probability.} In one 
instance\footnote{The case of Commitment with Symmetric Bidders 
described in Subsection
\ref{Subsection: Commitment With Symmetric Bidders} below.} it will
be necessary to consider explicitly an upper bound on bids, 
which will be denoted by $\bar{b}$ $>$ $\bar{v}$
for the first stage bidders.\footnote{A bound
$\bar{b}$ $=$ $\bar{v}$ does not pose a problem. If indeed 
$\bar{b}$ $=$ $\bar{v}$ then, for obvious reasons, 
(\ref{equation: b > v in general})
of Proposition \ref{proposition: a general overbidding result} below
only holds for $v \in (0,\bar{v})$ instead of any $v \in (0,\bar{v}]$.}
In all
other cases this is immaterial just as it is in virtually 
all existing auction 
models.

The (identical) bidding functions
of the first-stage bidders are denoted interchangeably 
by $b_i(\cdot)$ or $b(\cdot)$ while 
the bidding function
for bidder $N+1$ is denoted by $b_{N+1}(\cdot)$.

Finally, we refer to the case in which  {\it all}
$N+1$ bidders with the values as described above bid simultaneously
in a second price auction as the standard one-shot case. 

\subsection{No Commitment}\label{Subsection: No Commitment}
This is our point of departure. In the setup we just described we consider a two-stage auction as follows. 
At $T$ $=$ $1$ the $N$ first-stage bidders submit sealed bids. The highest bidder $i^*$ wins and pays a
price equal to second highest bid.

Bidder $i^*$ wins the right compete against $N+1$ in the second stage of the auction 
held at $T$ $=$ $2$. In the second stage auction $i^*$ and $N+1$ submit sealed 
bids.\footnote{Bidder $i^*$ submits a fresh bid in the second stage.} 
The highest bidder wins the object and 
pays a price equal
to the second highest bid --- the bid of the only other second-stage participant.

\subsection{Commitment With Symmetric 
Bidders}\label{Subsection: Commitment With Symmetric Bidders}
If the first stage bidders are committed to their bids 
in the sense that they cannot be changed in the 
second stage the picture changes considerably and several 
details begin to matter. 

The first case we consider is that of 
symmetric bidders in the sense that the value of $N+1$ 
has the same distribution as the first $N$ bidders. So, $G$ $=$ $F$. 

The highest bidder of the first stage $i^*$ goes
on to compete with bidder $N+1$ in the second stage
without the possibility of changing the bid. 
In the second stage the highest bidder wins 
and pays a price equal to the bid of the other bidder.

\subsection{Commitment With Asymmetric 
Bidders}\label{Subsection: Commitment With Asymmetric Bidders}
The auction
procedure is exactly the same as in subsection 
\ref{Subsection: Commitment With Symmetric Bidders}.
The first $N$ bidders are ex-ante identical just as in 
the previous cases. In this case bidder
$N+1$ is ``stronger'' in the sense that $E(w)$ $>$ $\bar{v}$.

\section{Results}\label{Section: Results}

\subsection{Equilibrium Behavior of 
Bidder $N+1$}\label{Subsection: Equilibrium Behavior of Bidder N+1}
It is clear from the description of our two-stage auction 
from the pont of view of bidder $N+1$ the set up always corresponds to a second price auction between
$N+1$ and the first stage winner $i^*$.

By completely standard arguments we then know that ``bidding his value'' 
is a weakly dominant strategy
for $N+1$. Hence, from now on we restrict attention to equilibria 
in which $N+1$ bids according to 
to this weakly dominant strategy. From now on
we assume that in equilibrium $b_{N+1}(w)$ $=$ $w$.

\subsection{Equilibrium in the No Commitment 
Case}\label{subsection: Equilibrium in the No Commitment Case}
Equilibrium for the model described in Subsection 
\ref{Subsection: No Commitment} is 
relatively straightforward to characterize.
\begin{proposition}[Bidding Functions In The 
No Commitment Case]\label{proposition: Bidding 
Functions In The No Commitment Case}
The following constitute an equilibrium in the case of no commitment.
\begin{eqnarray}\label{equation: Bidding Functions In The No Commitment Case}
b_i^I(v_i) \; = \; \int_{0}^{v_i} (v_i-\xi) g(\xi) \,d\xi \quad {\rm and} \quad b_i^{II}(v_i) = v_i,
\end{eqnarray}
where $b^I_i(\cdot)$ and $b^{II}_i(\cdot)$ are respectively the first and second stage (if he wins the first stage) 
bids for $i$.
And for bidder $N+1$,
\begin{eqnarray}\label{equation: N+1 
bidding function in the no commitment case}
\quad b_{N+1}(w) \; = \; w.
\end{eqnarray}
\end{proposition}

\begin{remark}[Value of Second 
Stage Participation]\label{remark: Value of Second Stage Participation}
By inspection of (\ref{equation: Bidding Functions In The No Commitment Case}) 
it is clear
that the $N$ first stage bidders simply bid ``truthfully'' their value of participating in 
the second stage of the
auction.
\end{remark}

\begin{remark}[Revenue Equivalence with No 
Commitment]\label{remark: Revenue Equivalence 
in the No commitment Case}
By inspection of (\ref{equation: Bidding Functions In The No Commitment Case}) 
and (\ref{equation: N+1 bidding function in the no commitment case}) it is clear
that the outcome of the two-stage auction without commitment 
is ex-post efficient. Therefore the Revenue Equivalence
Theorem implies that the expected revenue in this case 
is the same as the expected revenue
in the standard one-shot second price auction case.
\end{remark}

\begin{proposition}[Expected Revenue in the Standard and 
No Commitment Case]\label{proposition: Expected Revenue in the Standard and 
No Commitment Case}
Let $R^{NC}(N)$ be the expected revenue in the standard and in the no commitment case as a function of
the number $N$ of first stage bidders. Then\footnote{\rm 
Throughout, given a vector of random variables $x$ $=$ $(x_1, \ldots, x_L)$ of length $L$, we denote by 
$(x_1,...,x_L)_{(q:L)}$ the $q$-th (descending) order statistic of $x$.}
\begin{eqnarray}\label{equation: Expected Revenue in the Standard and 
No Commitment Case}
R^{NC}(N) \; = \; E[(v_1,...,v_N,w)_{(2:N+1)}]
\end{eqnarray}

When the values for all $N+1$ bidders are uniformly distributed on $[0,1]$, equation
(\ref{equation: Expected Revenue in the Standard and 
No Commitment Case}) becomes
\begin{eqnarray}\label{equation: Expected Revenue in the Standard and 
No Commitment Case with uniform values}
R^{NC}(N) \; = \; \frac{N}{N+2}.
\end{eqnarray}

\end{proposition}

\subsection{Equilibrium Overbidding With 
Commitment}\label{subsection: Equilibrium Overbidding With Commitment}

The commitment case is substantially 
different way from the no-commitment one and 
hence the model differs from the standard one-shot second price auction case.

Recall that in this case the first $N$ bidders bid {\it only once}. 
The highest bidder among the first $N$,
denoted $i^*$ goes on to compete with $N+1$ without revising his bid. 
A second price contest with $N+1$ then
decides who win the object and the price paid.

The two-stage structure of the auction generates increased 
competition in the first stage. 
Because of commitment, the only possibility to achieve a 
positive payoff for the first-stage bidders
is to commit to an above-value bid in order to attempt to 
win the right to compete in the second stage.
Formally, we prove the following.

\begin{proposition}[Above Value Bidding with 
Commitment]\label{proposition: a general overbidding result}
Consider distributions $F$ and $G$ as in Subsection \ref{Subsection: General} 
and the model with
commitment and either symmetric or asymmetric bidders.

Suppose that the (non-decreasing) bidding function $b(\cdot)$ 
for bidders $i=1,...,N$ 
induces an equilibrium in conjunction with $b_{N+1}(w) = w$. 
Then, for every $v \in (0,\bar{v}]$, we must have 
\begin{eqnarray}\label{equation: b > v in general}
b(v)\; > \; v
\end{eqnarray}
\end{proposition}

Proposition \ref{proposition: a general overbidding result} establishes 
that the two-stage
structure of the auction in the case of commitment generates 
bids that are above value for the first
$N$ bidders allowing for general distributions of values satisfying 
the conditions spelled out in 
Subsection \ref{Section: Setup}. Further characterization of the 
bidding functions and 
of the expected revenue will differ in crucial respects in the two 
cases of symmetric and asymmetric bidders
described above in Subsections \ref{Subsection: Commitment With 
Symmetric Bidders} and 
\ref{Subsection: Commitment With Asymmetric Bidders} respectively.

Restricting attention to values that are uniformly distributed 
yields further 
worthwhile characterizations. When we say that 
values are uniformly distributed
we mean that $F$ is the uniform distribution on $[0,1]$ and that $G$ 
is the uniform distribution
on $[0,\bar{w}]$ with $\bar{w}$ $\geq$ $1$.

Before proceeding further, we pause to notice that whenever 
Proposition \ref{proposition: a general overbidding result}
holds the Revenue Equivalence Theorem cannot be invoked to
pin down the expected revenue to the auctioneer.

\begin{remark}[Failure of Revenue Equivalence with 
Commitment]\label{remark: Revenue Equivalence 
Failure with Commitment and 
Symmetric Bidders} 
Combining the fact that bidder $N+1$ always bids his value truthfully
with the overbidding highlighted in Proposition 
\ref{proposition: a general overbidding result} it is clear that
in all the cases we consider where there is commitment
the outcome of the auction
may be ex-post inefficient with the object failing to be allocated
to the bidder with the highest value.

In all these cases the Revenue Equivalence
Theorem does {\rm not} pin down the expected revenue to the auctioneer
to be the same as in the standard one-shot second-price auction as
was the case for the no commitment model.

Clearly, the failure of the Revenue Equivalence Theorem 
leaves open the possibility that the revenue with commitment
may go up as well as down relative to the standard case.
\end{remark}

\subsection{Equilibrium with Commitment and Symmetric 
Bidders}\label{Subsection: Equilibrium with Commitment and Symmetric Bidders}
We are now ready
to characterize further 
the equilibrium of the model described in 
Subsection \ref{Subsection: Commitment With Symmetric Bidders} 
above.

\begin{proposition}[Commitment, 
Symmetric Bidders and Uniform 
Values]\label{proposition: Bidding Functions with Commitment and Symmetric Bidders}
Consider the case of Commitment, Symmetric Bidders and 
Uniform Values. In other words
assume that the values for all
$N+1$ bidders are uniformly distributed on $[0,1]$. Bids are 
restricted to be in $[0,\bar{b}]$ with
$\bar{b}$ $\geq$ $\bar{v}$ $=$ $\bar{w}$.

Then there exists a $\hat{v}_N$ $\in$ $(1/2, (N+1)/2N)$ 
such that the following constitute an equilibrium.
\begin{eqnarray}\label{equation: Bidding Functions with Commitment and Symmetric Bidders}
b_i(v_i) \; = \; \left\{ 
\begin{array}{lr}
    \displaystyle{\frac{2N}{N+1}}\; v_i & {\rm if} \;\; v_i\in[0,\hat{v}_N] \\
    \; & \; \\
    \bar{b} \quad & {\rm if} \;\; v_i \in (\hat{v}_N,1]
\end{array}\right. 
\qquad {\rm and} \quad b_{N+1}(w) \; = \; w
\end{eqnarray}
\end{proposition}

As expected,
equation (\ref{equation: Bidding Functions with Commitment and Symmetric Bidders}) 
confirms that in equilibrium
the first $N$ bidders bid above their values. Indeed from 
(\ref{equation: Bidding Functions with Commitment and Symmetric Bidders}) 
we observe that
they all bid $\bar{b}$ --- the maximum allowed --- whenever their value 
exceeds the cutoff value
$\hat{v}_N$. 
While the cutoff value $\hat{v}_N$ 
depends on $N$, the bid always jumps to the same value $\bar{b}$
after the cutoff is exceeded. 

As we noted in
Subsection \ref{Subsection: Equilibrium Behavior of Bidder N+1}, 
bidder $N+1$ always bids his value in
equilibrium. So in the case of symmetric bidders considered here, 
the bid of $N+1$ will never exceed $1$.
This in turn means that the first $N$ bidders have ``nothing to lose''
in bidding arbitrarily high once their bid exceeds $1$. 
Because the context against $N+1$ 
is a second price one, bidding any amount above $1$ 
will not change the price paid. This coupled 
with the general overbidding feature of 
Proposition \ref{proposition: a general overbidding result}
generates the need for the explicit bidding cap that we use here.

Our next concern is the effect on the auctioneer's 
expected revenue of the equilibrium
overbidding in this case.
The expected revenue falls below the standard case and 
can be characterized as follows.
\begin{proposition}[Revenue with Commitment and 
Symmetric Bidders]\label{proposition: Expected Revenue with Commitment and 
Symmetric Bidders}
Let $R^{CS}(N)$ be the expected revenue for the case of 
commitment, symmetric 
bidders and uniformly distributed values on $[0,1]$. 
Then 
\begin{eqnarray}\label{equation: revenue with symmestric and uniform}
R^{CS}(N) \; < \; 1/2 \; \leq \; R^{NC}(N) \qquad 
\forall \; N\; \geq \; 2
\end{eqnarray}

In the case of a general distribution of values 
$F$ $=$ $G$ our conclusion is limited to large
values of $N$.
\begin{eqnarray}\label{equation: limit revenue with symmetric and general}
\lim_{N\rightarrow \infty} R^{CS}(N) \; < \;  
\lim_{N\rightarrow \infty} R^{NC}(N)
\end{eqnarray}
\end{proposition}

\subsection{Equilibrium with Commitment and Asymmetric 
Bidders}\label{Subsection: Equilibrium with Commitment 
and Asymmetric Bidders}

We now turn to the model with Commitment and Asymmetric bidders 
described in Subsection
\ref{Subsection: Commitment With Asymmetric Bidders}. 
Bidder $N+1$ is stronger in the sense that
$F$ and $G$ are such that $E(w)$ $>$ $\bar{v}$.
Proposition \ref{proposition: a general overbidding result}
still applies in this case; the first $N$ bidders bid above their value.
The fact that $N+1$ is sufficiently stronger guarantees that
overbidding always has a potential cost for the first $N$ bidders. 
This in turn means that in this case there is no need to consider
explicitly a bidding cap $\bar{b}$ as we did
for the symmetric bidders case.

\begin{proposition}[Commitment, 
Asymmetric Bidders and Uniform Values]\label{proposition: Bidding Functions 
with Commitment and Asymmetric Bidders}

Assume that the bidders are asymmetric with $N+1$
being sufficiently stronger than the first 
$N$. Assume further that all 
values are uniformly distributed, the first $N$ on $[0,1]$. 

Then the following constitute an equilibrium.
\begin{eqnarray}\label{equation: Bidding Functions 
with Commitment and Asymmetric Bidders}
b_i(v_i) \; = \; \frac{2N}{N+1}v_i 
\quad {\rm and} \quad b_{N+1}(w) \; = \; w
\end{eqnarray}
\end{proposition}

The bidding above value generated 
by the two-stage structure of the auction with commintment
has a very different and ``unexpected'' 
effect in the case of asymmetric 
bidders. As we noted in Remark 
\ref{remark: Revenue Equivalence 
Failure with Commitment and 
Symmetric Bidders} since the outcome fails to be ex-post efficient
in this case, we cannot appeal to the Revenue Equivalence Theorem
to pin down the auctioneer's expected revenue relative to
the standard one-shot second-price auction case.

\begin{proposition}[Revenue with Commitment and 
Asymmetric Bidders]\label{proposition: Expected Revenue with Commitment and 
Asymmetric Bidders}
Let $R^{CA}(N)$ be the expected revenue for the case of 
commitment and asymmetric bidders
with bidder $N+1$ stronger than the first $N$ so that $E(w)$ $>$ $\bar{v}$. Assume general distributions
$F$ and $G$ for the bidders' values as in Subsection \ref{Subsection: General}. Assume further that
the bidding functions for the first $N$ bidders are 
differentiable.\footnote{\rm From Proposition 
\ref{proposition: Bidding Functions with Commitment and 
Asymmetric Bidders} we know this to be the case for uniformly distributed
values on $[0,1]$.}
Then, 
\begin{eqnarray}\label{equation: higher limit 
revenue with commitment and asymmetric bidders}
\lim_{N\rightarrow \infty} R^{CA}(N) \; > 
\; \lim_{N\rightarrow \infty} R^{NC}(N)
\end{eqnarray}
\end{proposition}

For large $N$ 
the two-stage structure of the auction yields an {\it increase}
in the auctioneer's expected revenue relative to the standard one-shot
second-price case. We believe this to be ``unexpected'' in the 
following sense.

Imagine adding a {\it single} bidder $N+1$ that is stronger than the first
$N$ to the standard one-shot second-price 
auction. This clearly has a vanishingly small effect on revenue. Bidder $N+1$ is 
indeed stronger, but of course he pays a price equal to the second 
highest bid. As $N$ becomes large, the second highest
value of the $N+1$ bidders, is almost surely arbitrarily close to (and just below)
$\bar{v}$. Hence bidder $N+1$ in the limit has no effect on 
revenue.



\newpage{}

\fontsize{12}{13.89}\selectfont
\bibliographystyle{ectranew}
\bibliography{large}

\newpage{}

\begin{appendix}

\proofof{Proposition \ref{proposition: Bidding Functions 
In The No Commitment Case}}
Equation (\ref{equation: N+1 
bidding function in the no commitment case}) follows from our remarks
about the bidding behavior of $N+1$ in Subsection 
\ref{Subsection: Equilibrium Behavior of Bidder N+1}).

Taking this as given, the expected benefit from participating in 
the second stage for a bidder $i=1,...,N$ with value 
$v_i \in [0,\bar{v}]$ is given by
\begin{eqnarray}\label{eqn app: benefit of secon stage with no commitment}
\theta(v_i) \equiv \int_{0}^{v_i} (v_i-\xi) g(\xi) \,d\xi
\end{eqnarray}

Observe that the first stage auction is equivalent to a standard one-shot 
second price auction in which bidders have values $\{\theta(v_i)\}_{i=1}^N$. By standard arguments bidding one's
own value is a weakly dominant strategy. Hence (\ref{equation: Bidding Functions In The No Commitment Case}) follows immediately. 
\qrendproof

\proofof{Proposition \ref{proposition: Expected Revenue in the Standard and 
No Commitment Case}}
Equation (\ref{equation: Expected Revenue in the Standard and 
No Commitment Case}) follows from the observation that
revenue equivalence 
with respect to the standard one-shot second-price auction holds in this
case. 
Equation (\ref{equation: Expected Revenue in the Standard and 
No Commitment Case with uniform values}) is a standard result about 
order statistics with uniform distributions.
\qrendproof

\proofof{Proposition \ref{proposition: a general overbidding result}}
Since in some cases constructing an equilibrium requires imposing a 
bidding cap on the first $N$ bidders
we let $\bar{b}$ $>$ $\bar{v}$ be such cap, with
the understanding that we may be in the case in which no such cap is 
imposed which here will correspond to setting $\bar{b}$ $=$
$\infty$.

Fix a bidding function for the first $N$ players forming an
equilibrium together with $b_{N+1}(w)$ $=$ $w$ as in the statement of
the proposition. 
Let $P(b_i;b(\cdot))$ be the probability that $i$ wins the 
first stage given bid $b_i$ and that all others bid according to $b(\cdot)$.

The expected payoff of a bidder $i=1,..,N$ with value $v_i$ 
in this equilibrium is then given by 
\begin{eqnarray}\label{i's interim payoff general overbidding proof}
P(b(v_i);b(\cdot)) \int_0^{b(v_i)} (v_i-\xi)g(\xi) \,d\xi. 
\end{eqnarray}
First note that we must have $P(b(v_i);b(\cdot))>0$ for 
all $v_i>0$ because $b(\cdot)$ is assumed to be 
non-decreasing ans ties are broken randomly.\footnote{See footnote
\ref{footnote: ties broken randomly}.}
Note also that the value of integral in 
(\ref{i's interim payoff general overbidding proof}) must also be 
positive otherwise $b(v_i)$ could not possibly be an optimal bid.
Hence, overall the expected payoff in 
(\ref{i's interim payoff general overbidding proof}) must be positive.

The remainder of the proof is in three steps.

\begin{step}\label{first step general overbidding}
$b(v_i)$ $\geq$ $v_i$ for all $v_i$ $\in$ $[0,\bar{v}]$
\end{step}
By way of contradiction suppose that $b(v_i)<v_i$ 
for some $v_i \in (0, \bar{v}]$. Recall that 
$i$'s expected payoff in equilibrium
conditional of having value $v_i$ is given by 
(\ref{i's interim payoff general overbidding proof}). 
Note that $P(b_i;b(\cdot))$
is non decreasing in $b_i$ and that since $b(v_i)<v_i$
\begin{eqnarray}\label{eqn app: partial of part of i's interim payoff}
\ds\frac{\partial}{\partial b_i} \, \ds\int_0^{b_i} (v_i-\xi) g(\xi) 
\, d\xi \;  = \; (v_i-b_i)\,g(b_i)\; > \; 0
\end{eqnarray}
Hence it follows that (without violating the bid cap, if any) 
$i$ can strictly increase his expected payoff by increasing his bid to
$b(v_i)+\epsilon$ for a sufficiently small $\epsilon>0$.

\begin{step}\label{step app: bid weakly greater than value}If 
$v_1<v_2$ and $b(v_2)<\bar{b}$, 
then $b(v_1)<b(v_2)$.\footnote{\rm In the case of no bid
cap $b(v_2)<\bar{b}$ is automatically satisfied.} 
\end{step}
If $v_1$ $=$ $0$ then it must clearly be that $b(v_1)$ $=$ $0$, and
if $v_2$ $>$ $0$ then $b(v_2)$ must be positive since the
interim payoff in (\ref{i's interim payoff general overbidding proof})
must be positive. So, there is nothing more to prove in this case.

Hence, using Step \ref{step app: bid weakly greater  than value}
it is sufficient to derive a contradiction from the case
$0$ $<$ $v_1$ $<$ $v_2$ and $b(v_1)=b(v_2)<\bar{b}$. 

Since $b(v_1)$ must be optimal given $v_1$ 
we must have for $\Delta$ $>$ $0$
\begin{eqnarray}\label{inequality for contradiction}
    P(b(v_1);b(\cdot))\int_0^{b(v_1)} (v_1-\xi)\, g(\xi) \,d\xi
    \; \geq 
    \; P(b(v_1)+\Delta;b(\cdot))\int_0^{b(v_1)+\Delta} (v_1-\xi) 
    \, g(\xi) \,d\xi.
\end{eqnarray}
where the left-hand side must be positive since $v_1$ $>$ $0$.

Furthermore, since by our contradiction
hypothesis $b(v_1)=b(v_2)$, it must be that
\begin{eqnarray}\label{eqn app: limit from above is greter because of tie}
    \lim_{\Delta \downarrow 0} P(b(v_1)+\Delta;b(\cdot)) > P(b(v_1);b(\cdot))
\end{eqnarray}
However (\ref{eqn app: limit from above is greter because of tie})
implies that as we take the limit for $\Delta$ $\downarrow$ $0$
inequality (\ref{inequality for contradiction}) must be false. 

\begin{step}\label{step app: final step in proof of general overbidding}
{\rm We can now conclude the proof of 
Proposition \ref{proposition: a general overbidding result}}.
\end{step}

Note that from Step \ref{step app: bid weakly greater than value} we know that
\begin{eqnarray}\label{eqn app: in step 3 we know step 2}
P(b(v_i), b(\cdot)) \; = \; [F(v_i)]^{N-1}
\end{eqnarray}

By way of contradiction suppose now that $b(v)=v$ for some $v\in(0,\bar{v}]$. Then, 
using Step \ref{step app: bid weakly greater than value} for any $\epsilon>0$, interim optimality implies
\begin{eqnarray}\nonumber
    F(v-\epsilon)^{N-1} \int_0^{b(v-\epsilon)} (v-\epsilon-\xi)g(\xi) \,d\xi \geq F(v)^{N-1} \int_0^{v} (v-\epsilon-\xi)g(\xi) \, d\xi
\end{eqnarray}
which we can rearrange as
\begin{eqnarray}
    \Big[F(v-\epsilon)^{N-1} - F(v)^{N-1}\Big] \int_0^{b(v-\epsilon)} (v-\epsilon-\xi)g(\xi) \,d\xi \geq F(v)^{N-1} \int_{b(v-\epsilon)}^{v} (v-\epsilon-\xi)g(\xi) \, d\xi
\end{eqnarray}

Since by Step \ref{first step general overbidding} we know that $b(v-\epsilon) \geq v-\epsilon$ (Step 1), we now must have
\begin{eqnarray}
\begin{array}{c}
    \left[F(v-\epsilon)^{N-1} - F(v)^{N-1}\right] \int_0^{b(v-\epsilon)} (v-\epsilon-\xi)g(\xi) \,d\xi \geq F(v)^{N-1} \int_{v-\epsilon}^{v} (v-\epsilon-\xi)g(\xi) \, d\xi \\ 
    \\
    = F(v)^{N-1} \left[ \int_{v-\epsilon}^{v} (v-\xi)g(\xi) \, d\xi - \epsilon [G(v)-G(v-\epsilon)] \right]
\end{array}
\end{eqnarray}
Hence we obtain that the following inequality must hold
\begin{eqnarray}\label{eqn app: last but one in step 3 of general overbidding}
\begin{array}{c}
    \displaystyle\frac{F(v-\epsilon)^{N-1} - F(v)^{N-1}}{\epsilon} \int_0^{b(v-\epsilon)} (v-\epsilon-\xi)g(\xi) \,d\xi \\ 
    \\
    \geq F(v)^{N-1} \left[ \frac{\int_{v-\epsilon}^{v} (v-\xi)g(\xi) \, d\xi}{\epsilon} - [G(v)-G(v-\epsilon)] \right]
\end{array}
\end{eqnarray}
Taking the limit as $\epsilon \downarrow 0$ on both sides of (\ref{eqn app: last but one in step 3 of general overbidding}), we now
obtain a contradiction since
\begin{eqnarray}\nonumber
- (N-1)F(v)^{N-2}f(v) \int_0^{v} (v-\xi)g(\xi) \,d\xi <0
\end{eqnarray}

\hfill\qrendproof

\proofof{Proposition \ref{proposition: Bidding Functions 
with Commitment and Symmetric Bidders}}
We begin by defining the functions
\begin{eqnarray}\label{eqn app: defining P A and B functions in proof of commitment and asymmetric bidders}
\begin{array}{lcl}
    P_N(v) & \equiv & \displaystyle{\sum_{i=0}^{N-1}
    \genfrac(){0pt}{0}{N-1}{i} \;
    v^{N-1-i} (1-v)^i \frac{1}{1+i}} \\
    & &\\
    A_N(v) & \equiv & \displaystyle{v^{N+1} \frac{2N}{(N+1)^2}} \\
    & &\\
    B_N(v) & \equiv & \displaystyle{(v-\frac{1}{2})P_N(v)}
\end{array}
\end{eqnarray}
since 
\begin{eqnarray}\nonumber
\genfrac(){0pt}{0}{N-1}{i}\, \frac{1}{1+i}= \genfrac(){0pt}{0}{N}{i+1} \, \frac{1}{N}
\end{eqnarray}
we  then immediately obtain
\begin{eqnarray}
    P_N(v) = \frac{1}{N} \frac{1}{1-v} \,\left[\sum_{j=1}^N 
    \genfrac(){0pt}{0}{N}{j}\, 
    (1-v)^j v^{N-j} \right] = 
    \frac{1}{N} \frac{1-v^N}{1-v} = \frac{1}{N} \sum_{j=0}^{N-1} v^j.
\end{eqnarray}
Our candidate  $\hat{v}_N$ as in the statement of the proposition is 
\begin{eqnarray}\label{eqn app: defining v_N in open interval}
    \hat{v}_N = \min \left\{v \in \left(\frac{1}{2},\frac{N+1}{2N}\right) \;\;\; {\rm such} \; {\rm that}  \;\;\; A_N(v) = B_N(v) \right\}
\end{eqnarray}
To see that the quantity in (\ref{eqn app: defining v_N in open interval}) is actually well defined, note that
$A_N(1/2)>B_N(1/2)=0$ and 
\begin{eqnarray}\nonumber
B_N((N+1)/2N)=\frac{1}{2N}\cdot P_N((N+1)/2N)> \frac{1}{2N} \Big(\frac{N+1}{2N}\Big)^{N-1} = A_N((N+1)/2N)
\end{eqnarray}
So that we know that $\hat{v}_N$ as above it is in fact interior to the interval in (\ref{eqn app: defining v_N in open interval}),
as required.

Given that all other players bid according to $b_{-i}(\cdot)$ as in equation (\ref{equation: Bidding Functions with Commitment and Symmetric Bidders}) 
in the statement of the proposition, the interim expected payoff of a bidder $i=1,...,N$ with bid $b_i$ and value $v_i$ is

\begin{eqnarray} \label{payoffs cases jump}
\begin{array}{lcr}
\pi(b_i;v_i,b_{-i}(\cdot)) & = \; \left\{ 
\begin{array}{lcr}
     \displaystyle{\left(\frac{N+1}{2N} b_i\right)^{N-1} \cdot b_i \cdot \left(v_i - \frac{b_i}{2}\right)} & {\rm if} & \displaystyle{b_i \leq \frac{2N}{N+1} \hat{v}_N} \\
    & & \\
    \displaystyle{\hat{v}_N^{N-1} \cdot b_i \left(v_i-\frac{b_i}{2}\right)} & {\rm if} &  \displaystyle{\frac{2N}{N+1}\hat{v}_N < b_i < 1} \\
    & & \\
    \displaystyle{\hat{v}_N^{N-1} \left(v_i-\frac{1}{2}\right)} & {\rm if } &\displaystyle{1 \leq b_i < \bar{b} \;\; ({\rm void \;\; if} \;\;  \bar{b}=1)} \\
   & & \\ 
   \displaystyle{P(\hat{v}_N)\left(v_i-\frac{1}{2}\right)} & {\rm if} & \displaystyle{b_i = \bar{b}}
\end{array}
\right.
\end{array}
\end{eqnarray}

To see why (\ref{payoffs cases jump}) holds, notice that the first term in each line corresponds to the probability of 
winning the first stage. Also, notice that the second term in each of the top two lines corresponds to the probability of winning the second stage. 
Finally, the last term in each of the top two lines is the expected payoff conditional on winning the second stage.

We can proceed to establish the interim optimality of the bidding function $b_i(\cdot)$
explicited in (\ref{equation: Bidding Functions with Commitment and Symmetric Bidders}). Before proceeding to examine specific cases, 
we note that it will never be an interim best reply to choose bid $b_i \in [1,\bar{b})$ because bidding $b_i=\bar{b}$ strictly improves $i$'s 
payoff.\footnote{The bid range $[1,\bar{b})$ is obviously empty if $\bar{b}=1$, so it this is irrelevant in this case.} We consider two cases separately.

\begin{case}\label{case app: vi in [0,vhatN]}
{\rm $v_i \in [0,\hat{v}_N]$.}
\end{case}
We begin by examining the optimal bid among the range $b_i \in [0,1)$. First notice that
\begin{eqnarray} \label{equation: maxinequality}
    \sup_{b_i \in [0,1)} \left(\frac{N+1}{2N}b_i\right)^{N-1} \cdot b_i \cdot \left(v_i - \frac{b_i}{2}\right) \geq \sup_{b_i \in [0,1)} \pi(b_i;v_i,b_{-i}(\cdot)).
\end{eqnarray}
Hence, if there is some bid $b_i\in [0,1)$ such that $\pi(b_i;v_i,b_{-i}(\cdot))$ equals the supremum on the \textit{LHS} of (\ref{equation: maxinequality}), 
then this $b_i$ must attain the supremum for the \textit{RHS} as well. With this in mind, let us first consider the \textit{LHS} of (\ref{equation: maxinequality}). 
From its first derivative, it is clear that the supremum is obtained at $b_i(v_i) = v_i {2N}/({N+1})<1$. 
Then, notice that for all $v_i \in [0,\hat{v}_N]$, we have $b_i(v_i) \leq \hat{v}_N {2N}/({N+1})$ and hence the first line of (\ref{payoffs cases jump}) 
is the relevant one when computing $\pi(b_i(v_i);v_i,b_{-i}(\cdot))$. It then follows that this bid $b_i(v_i)$ attains the supremum for the righthand 
side of (\ref{equation: maxinequality}) as well. Hence $b_i(v_i)$ as in (\ref{equation: Bidding Functions with Commitment and Symmetric Bidders}) 
is the optimal bid among bids $b_i \in [0,1)$.

Our next step is to compare this bid $b_i(v_i)$ with $\bar{b}$. The expected payoff given bid $b_i(v_i)$ is $A_N(v_i)$, 
whereas the expected payoff given bid $\bar{b}$ is $P(\hat{v}_N)(v_i- {1}/{2})$. Notice that these expected payoffs are equal for $v_i = \hat{v}_N$. 
Observe next that for all $v_i \in [0,\hat{v}_N]$,
\begin{eqnarray}
    \frac{{\rm d}}{{\rm d} v_i} A_N(v_i) = v_i^N \frac{2N}{N+1} < \hat{v}_N^{N-1} < P_N(\hat{v}_N).
\end{eqnarray}
Hence it must be that $A_N(v_i)>P(\hat{v}_N)(v_i - {1}/{2})$ for all $v_i \in [0,\hat{v}_N)$. 
Thus, it is clear that for Case \ref{case app: vi in [0,vhatN]}, $b_i(\cdot)$ as in 
(\ref{equation: Bidding Functions with Commitment and Symmetric Bidders}) is indeed the interim best reply.

\begin{case}\label{case app: vi in (vhatN,1]}
{\rm $v_i \in (\hat{v}_N,1]$.}
\end{case}

We begin by considering the optimal bid $b_i$ in the range $[0,\hat{v}_N 2N/({N+1})]$. Taking the first 
derivative of the first line in (\ref{payoffs cases jump}), it is clear that the optimal bid is always $\hat{v}_N 2N/({N+1})$. Therefore
\begin{eqnarray}\nonumber
    \max_{b_i \in [0,\hat{v}_N 2N/({N+1})} \pi(b_i;v_i,b_{-i}(\cdot)) = \displaystyle\frac{2N}{N+1} \hat{v}_N^N(v_i-\frac{N}{N+1}\hat{v}_N)
\end{eqnarray}
We can then establish that
\begin{eqnarray}\nonumber
\begin{array}{rcr}
    \hat{A}_N(v_i) \;\; \equiv \;\; \displaystyle{\sup_{b_i \in [0,1)}} \pi(b_i;v_i,b_{-i}(\cdot)) & = & \\
    & & \\
    \displaystyle{\sup_{b_i \in [\frac{2N}{N+1}\hat{v}_N,1)}} \pi(b_i;v_i,b_{-i}(\cdot)) & = & \left\{
    \begin{array}{lcr}
        \frac{2N}{N+1} \hat{v}_N^N(v_i-\frac{N}{N+1}\hat{v}_N) & {\rm if} & v_i \in (\hat{v}_N, \displaystyle{\frac{2N}{N+1}\hat{v}_N}] \\
        & & \\
        \hat{v}_N^{N-1} \displaystyle{\frac{v_i^2}{2}} & {\rm if} & v_i\in(\displaystyle{\frac{2N}{N+1}\hat{v}_N},1]
    \end{array}
    \right.
    \end{array}
\end{eqnarray}
We then observe that
\begin{eqnarray}
\begin{array}{lcr}
    \displaystyle\frac{{\rm d}}{{\rm d} v_i} \hat{A}_N(v_i) & = & \left\{ 
    \begin{array}{lcr}
        \displaystyle{\frac{2N}{N+1}} \hat{v}_N^N & {\rm if} & v_i \in (\hat{v}_N,\displaystyle\frac{2N}{N+1}\hat{v}_N] \\
        & & \\
        \hat{v}_N^{N-1} v_i & {\rm if} &  v_i\in(\displaystyle\frac{2N}{N+1}\hat{v}_N,1]
    \end{array}
\right.
\end{array}
\end{eqnarray}
so that $\hat{A}_N(v_i)$ is continuously differentiable on $v_i \in (\hat{v}_N,1]$. In addition, notice that the expected payoff 
given bid $\bar{b}$ is $P(\hat{v}_N)(v_i-{1}/{2})$, and observe that this is equal to $\hat{A}_N(\hat{v}_N)$ 
when $v_i = \hat{v}_N$. Finally observe that
\begin{eqnarray}
    P(\hat{v}_N)>\frac{{\rm d}}{{\rm d}v_i} \hat{A}_N(v_i)
\end{eqnarray}
for all $v_i \in (\hat{v}_N,1]$. Therefore the expected payoff given bid $\bar{b}$ is strictly better than $\hat{A}_N(v_i)$ for all $v_i \in (\hat{v}_N,1]$. 
Hence it follows that $b_i(\cdot)$ as in (\ref{equation: Bidding Functions with Commitment and Symmetric Bidders})
is the interim best reply for Case \ref{case app: vi in (vhatN,1]} as well. 

This is clearly enough to prove the claim.
\qrendproof

\proofof{Proposition \ref{proposition: Expected Revenue with Commitment and 
Symmetric Bidders}}
Since $N+1$ bids truthfully and there are only two bidders in the 
second stage, the second-price nature of the 
contest immediately implies that 
the realized revenue must be weakly less than $w$ under 
for any distribution of values across all $N+1$ bidders.
In addition, for the case of uniformly 
distributed values on $[0,1]$, inspection of the equilibrium 
in Proposition \ref{proposition: Bidding Functions with Commitment 
and Symmetric Bidders} immediately reveals that with non-zero probability, 
the realized revenue must be \textit{strictly} less than $w$. Thus, for 
the case of uniformly distributed values on $[0,1]$, we must have 
$R^{CS}(N)<1/2$ since the expected value of a uniform distribution 
on $[0,1]$ equals $1/2$. The remainder of the proposition for the uniform 
case follows directly from Proposition \ref{proposition: Expected Revenue 
in the Standard and No Commitment Case}.

For the case of general 
distributions case, observe once again that regardless of values
across all bidders, the realized revenue must be weakly less than $w$. 

It follows that $R^{CS}(N) \leq E(w)$ for all $N$. 
Then, notice that given our assumptions on the distribution $F$, 
the second-order statistic $(v_1,...,v_N,w)_{2:N} \overset{\mathrm{a.s.}}
{\to} \bar{v} = \bar{w}$. 
Thus, $\lim_{N \to \infty} R^{NC}(N) = \bar{w} > E(w)$, from which 
(\ref{equation: limit revenue with symmetric and general}) follows directly.
\qrendproof

\proofof{Proposition \ref{proposition: Bidding Functions 
with Commitment and Asymmetric Bidders}}
Notice that 
since $E(w)$ $=$ $\bar{w}/2$, we have that $\bar{w}/2$ $>$ $\bar{v}$ 
$=$ $1$. Next, suppose that all bidders other than 
$i$ $\neq$ $N+1$ bid as in 
(\ref{equation: Bidding Functions with Commitment and Asymmetric Bidders}).
Then if $i$ $\neq$ $N+1$ 
has value $v_i$ and bids $b_i$ his expected payoff is
given by 

\begin{eqnarray}\label{eqn app: i's intermim payoff asymmetric
uniform}
    \pi[b_i;v_i,b_{-i}(\cdot)] = 
    \left\{
    \begin{array}{lr}
        \left[\ds\frac{N+1}{2N}b_i\right]^{N-1} \; \ds\frac{b_i}
        {\bar{w}} \left[v_i-\ds\frac{b_i}{2}\right] & {\rm if} \; b_i 
        \in \left[0,\ds\frac{2N}{N+1}\right] \\
        \; & \; \\
        \ds\frac{b_i}{\bar{w}} \left(v_i-\ds\frac{b_i}{2}\right) 
        & {\rm if} \; b_i \in \left(\ds\frac{2N}{N+1},\bar{w}\right] \\
        \; &  \; \\
        v_i - \ds\frac{\bar{w}}{2} & {\rm if} \; b_i \in (\bar{w},\infty)
    \end{array}
    \right.
\end{eqnarray}

To see why (\ref{eqn app: i's intermim payoff asymmetric
uniform}) holds observe that the first term of the 
top two lines is the probability that $i$ by bidding $b_i$ wins the
first stage of the tournament in the two corresponding cases, and in those
same two cases $v_i$ $-$ $b_i/2$ is $i$'s expected payoff if he wins the 
second stage.

Now consider maximizing the top row of 
(\ref{eqn app: i's intermim payoff asymmetric uniform}) by choice of
$b_i$ $\in$ $[0,\infty)$. From the first order conditions the maximum is
attained by setting $b_i(v_i)$ as in 
(\ref{equation: Bidding Functions with Commitment and Asymmetric Bidders}).
By inspection of the second and third row of 
(\ref{eqn app: i's intermim payoff asymmetric uniform}), it is then clear
that $i$'s overall expected payoff as in 
(\ref{eqn app: i's intermim payoff asymmetric uniform}) is 
maximized by setting $b_i(v_i)$ as in 
(\ref{equation: Bidding Functions with Commitment and Asymmetric Bidders}),
and this is sufficient to prove the claim.\qrendproof

\proofof{Proposition \ref{proposition: Expected Revenue with Commitment and 
Asymmetric Bidders}} The proof is divided into three separate steps. The second step in turn is divided into three substeps.

\begin{step}\label{dadaumpa}
{\rm We prove that $\ds\sup_{N\geq2} b(\overline{v},N)$ $<$ $\overline{w}$.}
\end{step}
By assumption $E(w)>\bar{v}$. Therefore exists $\epsilon>0$ such that
\begin{eqnarray}
    E(w|w\leq \bar{w}-\epsilon) = \frac{1}{G(\bar{w}-\epsilon)}\int_0^{\bar{w}-\epsilon}\xi g(\xi) \,d\xi > 
    E(w) - \int_{\bar{w}-\epsilon}^{\bar{w}}\xi g(\xi) \,d\xi > \bar{v}.
\end{eqnarray}
Then, choose any $N\geq 2$ and suppose by way of contradiction that $b(\bar{v};N) \geq \bar{w}-\epsilon$. 
Then, the expected payoff of a bidder $i=1,...,N$ with value 
$\bar{v}$ in equilibrium is $G(b(\bar{v};N))[\bar{v}-E(w|w\leq b(\bar{v};N))] \leq G(b(\bar{v};N))[\bar{v}-E(w|w\leq \bar{w}-\epsilon)]<0$. 
Since any bidder can achieve nonnegative payoffs for sure by bidding $b_i=0$, this contradicts interim optimality. Hence, the claim
made in Step \ref{dadaumpa} has been established.

\begin{step}\label{lallallaero}
{\rm If $M>N$, then $b(v,M)>b(v,N)$ for all $v \in (0,\bar{v}]$.}
\end{step}
This step is proved in three distinct substeps.

{\bf Step \ref{lallallaero}-a}: $b(\cdot,N)$ satisfies the ODE
\begin{eqnarray}\label{eqn app: ODE in step 2a}
\frac{\partial}{\partial b}b(v;N) = H^N(b(v;N),v) \quad {\rm for} \; {\rm all} \quad v \in (0,\bar{v}]
\end{eqnarray}
where
\begin{eqnarray}\label{eqn app: definition of H in ODE}
H^N(b,v) \equiv (N-1) \cdot \frac{1}{b-v} \cdot \frac{f(v)}{F(v)} \cdot \ds\frac{G(b)}{g(b)} \cdot E_w[v-w|w\leq b]
\end{eqnarray}
for all $(b,v) \in (0,\bar{w}]\times (0,\bar{v}]$ such that $b>v$. 

The claim follows directly from rearranging the FOC of the first $N$ bidders, and using Step \ref{dadaumpa}.

{\bf Step \ref{lallallaero}-b}: $b(0,N)$ $=$ $0$.

Denote
\begin{eqnarray}\nonumber
T(b,b(\cdot,N)) \equiv F(b^{-1}(b_i,N))^{N-1}
\quad {\rm
and} \quad
S(b_i,v_i) \equiv  \int_{0}^{b_i} (v_i-\xi) g(\xi) \,d\xi.
\end{eqnarray}
Then, let $v_i \in (0,\bar{v}]$. It immediately follows that $T(b(v_i,N),b(\cdot,N))=F(v_i)^{N-1}>0$. In addition, 
\begin{eqnarray}
\frac{\partial}{\partial b}T(b(v_i,N),b(\cdot;N)) = (N-1) F(v_i)^{N-2} f(v_i)/b(v_i,N)>0
\end{eqnarray}
Next, by way of contradiction suppose that $b(0,N)$ $>$ $0$. Then
\begin{eqnarray}
S(b(v_i,N),v_i) \; = \; G(b(v_i, N))\left[v_i-E_w[w | b(v_i, N)>w]\right] \leq G(b(v_i,N)) \left[v_i - E_w[w|b(0,N)>w] \right]
\end{eqnarray}

Now choose any $v_i \in (0,E_w[w | b(0,N)>w])$. It follows that $T(b(v_i,N),b(\cdot,N))>0$ and 
$S(b(v_i,N),v_i)<0$. Therefore $\pi(b(v_i,N), v_i,b_{-i}(\cdot,N))<0$. 
This contradicts interim optimality and hence establishes the claim.

{\bf Step \ref{lallallaero}-c}: Fix $M$ $>$ $N$. Then there exists a $v_\epsilon>0$ such that $b(v,M)>b(v,N)$ for all $v \in (0,v_\epsilon)$.

We begin by observing that using (\ref{eqn app: ODE in step 2a}) and (\ref{eqn app: definition of H in ODE}) 

\begin{eqnarray}\label{eqn app: this the famous A.*}
H^N(b,v) \; < \; H^M(b,v)  \;\; \forall \; b \in (v,b(v)], \;\; \forall \; v\in(0,\bar{v}]
\end{eqnarray}
moreover
\begin{eqnarray}\label{eqn app: partial of HN}
\frac{\partial}{\partial b}H^N(b,v) = (N-1) \frac{f(v)}{F(v)} \;  \cdot \; \frac{\partial}{\partial b} \left[ \frac{1}{b-v}\; \frac{1}{g(b)} \; \int_{0}^{b} (v-\xi) g(\xi) \,d\xi \right]
\end{eqnarray}
From (\ref{eqn app: partial of HN}) we get
\begin{eqnarray}
{\rm sign}\left\{\frac{\partial}{\partial b}H^N(b,v)\right\} = {\rm sign}\left\{\frac{\partial}{\partial b} \left[ \frac{1}{b-v} \frac{1}{g(b)} \int_{0}^{b} (v-\xi) g(\xi) \,d\xi \right]\right\}
\end{eqnarray}
and
\begin{eqnarray}
{\rm sign}\left\{\frac{\partial}{\partial b}H^N(b,v)\right\} = {\rm sign}
\left\{-1-\frac{1}{b-v}\frac{1}{g(b)}\left[\frac{g'(b)}{g(b)}+\frac{1}{b-v}\right] \int_{0}^{b} (v-\xi) g(\xi) \,d\xi \right\}
\end{eqnarray}
By assumption $g'$ is bounded at $0$ and $g(0)>0$. Therefore there exists a $b_\delta>0$ and $K<\infty$ such that
\begin{eqnarray}
\left | \ds\frac{g'(b)}{g(b)}\right| \; <K\; \qquad \forall \;  b \in [0,b_\delta)
\end{eqnarray}
From Step \ref{lallallaero}-b we know that that $\lim_{v \downarrow 0} b(v,N) - v = 0$. 
It then follows that there exists $v_\epsilon>0$ such that for all $v\in(0,v_\epsilon)$,
\begin{eqnarray}
\ds\frac{1}{b(v,N)-v} > K \qquad {\rm and} \qquad b(v,N) < b_\delta
\end{eqnarray}
A direct implication of interim optimality is that 
\begin{eqnarray}\nonumber
\ds\int_{0}^{b(v,N)} (v-\xi) g(\xi) \,d\xi>0 \qquad  \forall v>0 
\end{eqnarray}
Hence
\begin{eqnarray}\label{eqn app: this is another one that was no numbered}
\frac{\partial}{\partial b}H^N(b(v,N),v)<0 \qquad \forall \, v \in (0,v_\epsilon)
\end{eqnarray}
By way of contradiction now suppose that for every $v_{\epsilon}$ we van find a $v^*$ $\in$ $(0,v_{\epsilon})$ such that
\begin{eqnarray}\nonumber
b(v^*,M) \; \leq \; b(v^*,N)
\end{eqnarray}
Then, using (\ref{eqn app: ODE in step 2a}), (\ref{eqn app: definition of H in ODE}), (\ref{eqn app: this the famous A.*}) and
(\ref{eqn app: this is another one that was no numbered}), we get
\begin{eqnarray}\nonumber
\frac{\partial}{\partial v} b(v^*;N) = H^N(b(v^*;N),v^*) \leq H^N(b(v^*;M),v^*) < H^M(b(v^*;M),v^*) = \frac{\partial}{\partial v}b(v^*;M).
\end{eqnarray}
Hence there exists $\tilde{v}\in (0,v^*)$ arbitrarily close to $v^*$ such that
\begin{eqnarray}
b(\tilde{v},M)<b(\tilde{v},N)
\end{eqnarray}

Since $b(0;M)=b(0;N)=0,$ there exists $\hat{v} \in (0,\tilde{v})$ such that
\begin{eqnarray}
 \frac{\partial}{\partial v}b(\hat{v}M) <  \frac{\partial}{\partial v}b(\hat{v},N)
\end{eqnarray}
from which we have 
\begin{eqnarray}
H^N(b(\hat{v},M);\hat{v})<H^M(b(\hat{v},M);\hat{v})=\frac{\partial}{\partial v}b\hat{v},M)<\frac{\partial}{\partial v}b(\hat{v},N)=H^N(b(\hat{v},N),\hat{v})
\end{eqnarray} 
and hence
\begin{eqnarray}
b(\hat{v},M) > b(\hat{v},N)
\end{eqnarray}

Hence $v^{**} \equiv \sup \{v\in(0,\tilde{v}) | b(v;M)>b(v;N)\} \in (\hat{v},\tilde{v})$. Also, by continuity, $b(v^{**};M) = b(v^{**};N).$ 
However, this in turn implies that 
\begin{eqnarray}
\frac{\partial}{\partial v}b(v^{**},M)\; >\; \frac{\partial}{\partial v}b(v^{**},N)
\end{eqnarray}
This clearly contradicts the definition of $v^{**}$, and hence establishes the claim.

We can now conclude the proof of the claim in Step \ref{lallallaero}.

Firstly, from Step \ref{lallallaero}-c, there exists $v_\epsilon>0$ such that $b(v,M)>b(v,N)$ for all $v \in (0,v_\epsilon]$. 
By way of contradiction suppose that $\{v \in (v_\epsilon,\bar{v}] \, | \, b(v,M) \leq b(v;N)\}$ $\neq$ $\emptyset$. Then, denote 
$v^{***} \equiv \inf \{v \in (v_\epsilon,\bar{v}] \, | \,  b(v,M) \leq b(v,N)\}$. Since $b(v_\epsilon,M)>b(v_\epsilon,N)$, 
we must have $v^{***} \in (v_\epsilon,\bar{v}].$ Then, by continuity of $b(\cdot;N)$ and $b(\cdot;M)$, 
we must have $b(v^{***},N)=b(v^{***},M).$ However, this in turn implies 
\begin{eqnarray}
\frac{\partial}{\partial v}b(v^{***},M)\; > \; \frac{\partial}{\partial v}b(v^{***},N)
\end{eqnarray}
This contradicts the definition of $v^{***}$ and hence establishes the claim.

\begin{step}\label{lallallaerolallala}
{\rm This step simply concludes the proof of Proposition \ref{proposition: Expected Revenue with Commitment and 
Asymmetric Bidders}.}
\end{step}

From Steps \ref{dadaumpa} and \ref{lallallaero} we 
have that the limit $\bar{b}_\infty \equiv \lim_{N \to \infty} b(\bar{v};,N)$ is well-defined with $\bar{b}_\infty \in (\bar{v},\bar{w})$. 
A sharper characterization of $\bar{b}_\infty$ will allow us to complete the proof.

Define the value function
\begin{eqnarray}\label{eqn app: defintion of value W}
W^N(v) \equiv \pi^N(b(v,N),v,b_{-i}(\cdot,N)) = F(v)^{N-1} \int_0^{b(v,N)} (v-\xi) g(\xi) \,d\xi \quad \forall v \in [0,\bar{v}].
\end{eqnarray}
By the envelope theorem,
\begin{eqnarray}\nonumber
\frac{\partial}{\partial v} W^N(v) \; = \; F(v)^{N-1} \cdot G(b(v,N))
\end{eqnarray}
and therefore
\begin{eqnarray}\nonumber
W^N(\bar{v}) =  \int_{0}^{\bar{v}} F(v)^{N-1} \cdot G(b(v,N)) \,dv 
\end{eqnarray}
Then, by the dominated convergence theorem,
\begin{eqnarray}\nonumber
\lim_{N\to\infty} W^N(\bar{v}) = 0.
\end{eqnarray}
However, by definition (\ref{eqn app: defintion of value W}) we also have that
\begin{eqnarray}\nonumber
W^N(\bar{v}) = E_w\left[(b(\bar{v},N)\geq w) \cdot (\bar{v}-w)\right]
\end{eqnarray}
By the dominated convergence theorem again,
\begin{eqnarray}\nonumber
\lim_{N \to \infty} W^N(\bar{v}) = E_w\left[(\bar{b}_\infty \geq w)\cdot (\bar{v}-w)\right].
\end{eqnarray}
Hence
\begin{eqnarray}\nonumber
E_w[w\,|\,w \leq \bar{b}_\infty] \; = \; \bar{v}.
\end{eqnarray}

Denote as ${\cR}^{CA}(N)$ the random variable of the revenue from the tournament auction with commitment and $N$ bidders. 
Furthermore, denote as ${\cR}^{NC}(N)$ the random variable of the revenue from the one-shot second-price auction with $N$ bidders. 

Then, observe that
\begin{eqnarray}\nonumber
\mathcal{R}^{CA}(N) = \min \left\{b\left(\max_{i=1,...,N} v_i,N\right),w\right\}\; \overset{\mathrm{a.s.}}{\rightarrow} \; \min\left\{\bar{b}_\infty,w\right\}
\end{eqnarray}

\begin{eqnarray}\label{eqn app: as convergence to vbar}
\mathcal{R}^{NC}(N) \; \overset{\mathrm{a.s.}}{\rightarrow} \;  \bar{v}
\end{eqnarray}
Using once more the monotone convergence theorem, it then follows that
\begin{eqnarray}
\begin{array}{c}
\ds\lim_{N \to \infty} E\left[\mathcal{R}^{CA}(N)\right] \; = \; E\left[\min\{\bar{b}_\infty,w\}\right] \; = \\
\\
\left[1-G(\bar{b}_\infty)]\right]\bar{b}_\infty + G(\bar{b}_{\infty}) E_w\left[w \, | \, w \leq \bar{b}_\infty\right] \;  > \; \bar{v}
\end{array}
\end{eqnarray}
which, using (\ref{eqn app: as convergence to vbar}), is clearly enough to complete the proof. \qrendproof
\end{appendix}
\end{document}